# Comparison of volatility distributions in the periods of booms and stagnations: an empirical study on stock price indices

Taisei Kaizoji

Econophysics Laboratory, Division of Social Sciences,
International Christian University
Osawa, Mitaka, Tokyo 181-0011 Japan.
e-mail: kaizoji@icu.ac.jp,
homepage: http://subsite.icu.ac.jp/people/kaizoji/

**Abstract**

The aim of this paper is to compare statistical properties of stock price indices in periods of booms with those in periods of stagnations. We use the daily data of the four stock price indices in the major stock markets in the world: (i) the Nikkei 225 index (Nikkei 225) from January 4, 1975 to August 18, 2004, of (ii) the Dow Jones Industrial Average (DJIA) from January 2, 1946 to August 18, 2004, of (iii) Standard and Poor's 500 index (SP500) from November 22, 1982 to August 18, 2004, and of (iii) the Financial Times Stock Exchange 100 index (FT 100) from April 2, 1984 to August 18, 2004. We divide the time series of each of these indices in the two periods: booms and stagnations, and investigate the statistical properties of *absolute log returns*, which is a typical measure of volatility, for each period. We find that (i) the tail of the distribution of the absolute log-returns is approximated by a *power-law* function with the exponent close to 3 in the periods of booms while the distribution is described by an *exponential* function with the scale parameter close to unity in the periods of stagnations.

## 1. Introduction

The statistical properties of the fluctuations of financial prices have been widely researched since Mandelbrot (1963) and Fama (1965) presented evidence that return distributions can be well described by a symmetric Levy stable law with tail index close to 1.7. In particular, a large number of empirical studies have shown that the tails of the distributions of returns and volatility follow approximately a power law with estimates of the tail index falling in the range 2 to 4 for large value of returns and volatility. (See, for examples, de Vries (1994); Pagan (1996); Longin (1996), Lux (1996); Guillaume et al. (1997); Muller et al. (1998); Gopikrishnan et al. (1998), Gopikrishnan et al. (1999), Plerou et al. (1999), Liu et al. (1999)). However, there is evidence against power-law tails too. For instance, Barndorff-Nielsen (1997), Eberlein et al. (1998) have respectively fitted the distributions of returns using normal inverse Gaussian, and hyperbolic distribution. Laherrere and Sornette (1999) have suggested to describe the distributions of returns by the Stretched-Exponential distribution. Dragulescu and Yakovenko (2002), and Kaizoji and Kaizoji (2003) have shown that the distributions of returns have been approximated by exponential distributions. More recently, Malevergne, Pisarenko and Sornette (2004) have suggested that the tails ultimately decay slower than any stretched exponential distribution but probably faster than power laws with reasonable exponents as a result from various statistical tests of returns.

Thus opinions vary among scientists as to the shape of the tail of the distribution of returns (and volatility). While there is fairly general agreement that the distribution of returns and volatility has power-like tail for large values of returns and volatility, there is still room for a considerable measure of disagreement about the hypothesis. At the moment we can only say with fair certainty that (i) the power-law tail of the distribution of returns and volatility is not an universal law and (ii) the tails of the distribution of returns and volatility are heavier than a Gaussian, and are between power-law and exponential.

There is one other thing that is important for understanding of price movements in financial markets. It is a fact that the financial market has repeated booms (or bull market) and stagnations (or bear market). To ignore this fact is to miss the reason why price fluctuations are caused. This is an important fact to stress. However, in a large number of empirical studies, which have been made on statistical properties of returns and volatility in financial markets, little attention has been given to the relationship between market situations and price fluctuations. Kaizoji (2004a) investigates this subject using the historical data of the Nikkei 225 index. We found that the shape of the volatility distribution in the period of booms was different from that in the period of stagnations. The purpose of this paper is to examine further the statistical properties of volatility distribution from this viewpoint. We use the daily data of the four stock price indices of the three major stock markets in the world: the Nikkei 225 index, the DJIA. SP500, and FT100, and compare the shape of the volatility distribution for each of the stock price indices in the periods of booms with that in the period of stagnations. We find that (i) the tail of the distribution of the absolute log-returns is approximated by a power-law function with the exponent close to 3 in the periods of booms while the distribution is described by an exponential function with the scale parameter close to unity in the periods of stagnations. These indicate that so far as the stock price indices we used are concerned, the same observation on the volatility distribution holds in all cases.

The rest of the paper is organized as follows: the next section analyzes the stock price indices and shows the empirical findings. Section 3 gives concluding remarks.

## 2. Empirical analysis

### 2.1. Stock Price Indices

We investigate quantitatively the four stock price indices[1]of the three major stock markets in the world, that is, (a) the Nikkei 225 index (Nikkei 225), which is the price-weighted average of the stock prices for 225 large companies listed in the Tokyo Stock Exchange, (b) the Dow Jones Industrial Average (DJIA) which is the price-weighted average of 30 significant stocks traded on the New York Stock Exchange and Nasdaq, (c) Standard and Proor's 500 index (SP 500) which is a market-value weighted index of 500 stocks chosen for market size, liquidity, and industry group representation, and (d) FT 100, which is similar to SP 500, and a market-value weighted index of shares of the top 100 UK companies ranked by market capitalization. Figure 1(a)-(d) show the daily series of the four stock price indices: (a) the Nikkei 225 from January 4, 1975 to August 18, 2004, (b) DJIA from January 2, 1946 to August 18, 2004, (c) SP 500 from November 22, 1982 to August 18, 2004, and (d) FT 100 from April 2, 1984 to August 18, 2004.

[1] The prices of the indices are close prices which are adjusted for dividends and splits.

After booms of a long period of time, the Nikkei 225 reached a high of almost 40,000 yen on the last trading day of the decade of the 1980s, and then from the beginning trading day of 1990 to mid-August 1992, the index had declined to 14,309, a drop of about 63 percent. A prolonged stagnation of the Japanese stock market started from the beginning of 1990.

The time series of the DJIA and SP500 had the apparent positive trends until the beginning of 2000. Particularly these indices surged from the mid-1990s. There is no doubt that this stock market booms in history were propelled by the phenomenal growth of the Internet which has added a whole new stratum of industry to the American economy. However, the stock market booms in the US stock markets collapsed at the beginning of 2000, and the descent of the US markets started. The DJIA peaked at 11722.98 on January 14, 2000, and dropped to 7286.27 on October 9, 2002 by 38 percent. SP500 arrived at peak for 1527.46 on March 24, 2000 and hit the bottom for 776.76 on October 10, 2002. SP500 dropped by 50 percent. Similarly FT100 reached a high of 6930.2 on December 30, 2000 and the descent started from the time. FT100 dropped to 3287 on March 12, 2003 by 53 percent.

From these observations we divide the time series of these indices in the two periods on the day of the highest value. We define the period a period until it reaches the highest value as the period of booms and the period after that as stagnations, respectively. The periods of booms and stagnations for each index of the four indices are collected into Table 1.

| Name of Index | The Period of Booms | The Period of Stagnations |
|---|---|---|
| Nikkei225 | Jan. 4, 1975 – Dec. 30, 1989 | Jan. 4, 1990 -Aug.18, 2004 |
| DJIA | Jan. 2, 1946 – Jan. 14, 2000 | Jan. 18, 2000-Aug. 18, 2004 |
| SP500 | Nov. 22, 1982-Mar. 24, 2000 | Mar. 27, 2000-Aug. 18, 2004 |
| FT100 | Mar. 3, 1984-Dec. 30, 1999 | Jan. 4, 2000-Aug. 18, 2004 |

Table 1: The periods of booms and stagnations.

### 2.2. Comparisons of the distributions of absolute log returns

In this paper we investigate the shape of distributions of absolute log returns of the stock price indices. We concentrate to compare the shape of the distribution of volatility in the period of booms with that in the period of stagnations. We use absolute log return, which is a typical measure of volatility. The absolute log returns is defined as $|R(t)| = |\ln S(t) - \ln S(t-1)|$, where $S(t)$ denotes the index at the date $t$. We normalize the absolute log-return $|R(t)|$ using the standard deviation. The normalized absolute log return $V(t)$ is defined as $V(t) = |R(t)| / \sigma$ where $\sigma$ denotes the standard deviation of $|R(t)|$.

Figure 2 (a)-(d) show the semi-log plot of the complementary cumulative distribution function of the normalized absolute log-returns $V$ for each of the four stock price indices.

Each panel compares the distribution of *V* for an index in the period of booms with that in the period of stagnations. The circles represent the distribution in the period of booms and triangles that in the period of stagnations. In the all panels it follows that the tail of the volatility distribution of *V* is heavier in the period of booms than in the period of stagnations. We shall now look more carefully into the difference between the two distributions. To this aim, we attempt to fit the empirical distributions with the two specific distributions, that is, an exponential and power-law function below.

The panels (a)-(h) of Figure 3 show the semi-log plots of the complementary cumulative distribution of *V* for each of the four indices: Nikkei225, DJIA, SP500, and FT100 in the period of booms and that in the period of stagnations, respectively. The solid lines in all panels represent the fits of the exponential distribution,

$$P(V > x) \propto \exp(-\frac{x}{\beta}) \qquad (1)$$

where the scale parameter $\beta$ is estimated from the data using a least squared method. In all cases of the period of stagnations, which are panels (b), (d), (f) and (h), the exponential distribution (1) describes very well the distributions of *V* over a whole range of values of *V* except for only the tow extreme value of V, that is, . In panel (b) the Nikkei 225 had one extreme value that occurred on September 28, 1990 in the Japanese stock markets and that occurred on September 10, 2001 in US stock markets. The jump of Nikkei 225 perhaps was caused by investors' speculation on the 1990 Gulf War. The extreme value of DJIA was caused by terror attack in New York on September 10, 2001[2]. The scale parameter $\beta$ is estimated from the data except for these two extreme values using a least squared method is collected in Table 2. In all cases the values of the estimated $\beta$ are very close to unity.

| Name of Index | The scale parameter $\beta$ | $R^2$ |
|---|---|---|
| Nikkei 225 | 1.02 | 0.995 |
| DJIA | 1.09 | 0.995 |
| SP500 | 0.99 | 0.997 |
| FT100 | 0.99 | 0.999 |

Table 2: The scale parameter $\beta$ of an exponential function (1) estimated from the data using the least squared method. $R^2$ denotes the coefficient of determinant.

On the other hand the panels (a), (c), (e), and (g) of Figure 3 show the complementary cumulative distribution of *V* in the period of booms for each of the four indices in the semi-log plots. The solid lines in all panels represent the fits of the exponential distribution estimated from the data of only the low values of *V* using a least squared method. In these cases the low values of *V* are only approximately well described by the exponential distribution (1), but completely fails in describing the large values of *V*.

[2] The extreme value does not appear in SP500. We would like to note that this perhaps originate in a difference between the calculation methods of the DJIA and the SP 500. In DJIA Higher-priced stocks affect the average greater than lower-priced ones, while regardless of stock price, a percentage change will be reflected the same on the index in SP500.

Apparently, an exponential distribution underestimates large values of *V*.

The panels (a)-(d) of Figure 4 show the complementary cumulative distribution of V in the period of stagnations for each of the four indices in the log-log plots. The solid lines in all panels represent the best fits of the power-law distribution for the large values of *V*,

$$P(V > x) \propto x^{-\alpha} . \qquad (2)$$

The power-law exponent $\alpha$ is estimated from the data of the large values of *V* using the least squared method. The best fits succeed in describing approximately large values of *V*. Table 3 collect the power-law exponent $\alpha$ estimated. The values of the estimated $\alpha$ are in the range from 2.8 to 3.7.

| Name of Index | The power-law exponent $\alpha$ | $R^2$ |
|---|---|---|
| Nikkei 225 | 2.83 | 0.992 |
| DJIA | 3.69 | 0.995 |
| SP500 | 3.26 | 0.986 |
| FT100 | 3.16 | 0.986 |

Table 3: The power-law exponent $\alpha$ of a power-law function (2) estimated from the data using the least squared method. $R^2$ denotes the coefficient of determinant.

Finally The panels (a) and (b) of Figure 5 show the complementary cumulative distributions of *V* for the four indices in the period of booms in a semi-log scale, and those in the period of stagnations in a log-log scale. The two figures confirm that the shape of the fourth volatility distributions in the periods of booms and of stagnations is almost the same, respectively.

## 3. Concluding remarks

In this paper we focus on comparisons of shape of the distributions of absolute log returns in the period of booms with those in the period of stagnations for the four major stock price indices. We find that the complementary cumulative distribution in the period of booms is very well described by exponential distribution with the scale parameter close to unity while the complementary cumulative distribution in the large value of the absolute log returns is approximated by power-law distribution with the exponent in the range of 2.8 to 3.8. The latter is complete agreement with numerous evidences to show that the tail of the distribution of returns and volatility for large values of volatility follow approximately a power law with the estimates of the exponent $\alpha$ falling in the range 2 to 4. We are now able to see that the statistical properties of volatility for stock price index are changed according to situations of the stock markets. Our findings make it clear that we must look more carefully into the relationship between regimes of markets and volatility in order to fully understand price fluctuations in financial markets.

The question which we must consider next is the reasons why and how the differences are created. That traders' herd behavior may help account for it would be accepted by most people. Recently we have proposed a stochastic model (Kaizoji (2004b)) that may offer the key to an understanding of the empirical findings we present here. The results of the numerical simulation of the model suggest the following: in the period of booms, the noise traders' herd behavior strongly influences to the stock market and generate power-law tails of the volatility distribution while in the period of stagnations a large number of noise traders leave a stock market and interplay with the noise traders become weak, so that exponential tails of the volatility distribution is observed. However it remains an unsettled question what causes switching from boom to stagnation.

Our findings may provide a starting point to make a new tool of risk management of index fund in financial markets, but to apply the rule we show here to risk management, we need to establish the framework of analysis and refine the statistical methods. We began with a simple observation on the stock price indices, and divided the price series into the two periods: booms and stagnations. However, there is room for further investigation on how to split the price series into periods according to the situations of markets. It is also worth while examining the statistical tools to estimate the tail of the volatility distribution more closely and comprehensively. As Malevergne, Pisarenko, and Sornette (2004) have suggested, the log-Weibull model, which provides a smooth interpolation between exponential distribution and power-law distribution, will be considered as an appropriate approximation of the volatility distributions. These studies will be left for future work.

## 4. Acknowledgements

An earlier version of this paper was presented at the 8th Annual Workshop on Economics with Heterogeneous Interacting Agents (WEHIA2003) hold at Institute in World Economy, Kiel, Germany, May 29-31, 2003. My special thanks are due to Prof. Thomas Lux and Prof. Enrico Scalas for valuable comments for an earlier version. The writing of this revised version was made largely in the Max Planck Institute for research into Economic Systems at Jena, Germany. I wish to express my gratitude to Prof. Ulrich Witt and Prof. Thomas Brenner and the staff for inviting me to their institute and providing me with wonderful research environment.

Financial support by the Japan Society for the Promotion of Science under the Grant-in-Aid, No. 06632 is gratefully acknowledged. All remaining errors, of course, are mine.

## 5. References

Barndorff-Nielsen, O.E., 1997, Normal inverse Gaussian distributions and the modelling of stock returns, Scandinavian Journal of Statistics 24, 1-13.

Dacorogna M.M., U.A. Muller, O.V. Pictet and C.G. de Vries, 1992, The distribution of extremal foreign exchange rate returns in large date sets, Working Paper, Olsen and Associates Internal Documents UAM, 19921022.

Dragulescu, A.A. and V.M. Yakovenko, 2002, Probability distribution of returns for a model with stochastic volatility,

Eberlein, E., Keller, U. and Prause, K., 1998, New insights into smile, mispricing and value at risk: the hyperbolic model, Journal of Business 71, 371-405.

Embrechts P., C.P. Kl¨uppelberg and T. Mikosh, 1997, Modelling Extremal Events (Springer-Verlag).

Fama E.F., 1965, The Behavior of Stock Market Prices, Journal of Business 38, 34-105.

Gopikrishnan P., M. Meyer, L.A.N. Amaral and H.E. Stanley, 1998, Inverse Cubic Law for the, Distribution of Stock Price Variations, European Physical Journal B 3, 139-140.

Gopikrishnan, P., V. Plerou, L. A. N. Amaral, M. Meyer, and H. E. Stanley, 1999, Scaling of the distributions of fluctuations of financial market indices, Phys. Rev. E 60, 5305-5316.

Guillaume D.M., M.M. Dacorogna, R.R. Dav´e, U.A. Muller, R.B. Olsen and O.V. Pictet, 1997, From the bird's eye to the microscope: a survey of new stylized facts of the intra-day foreign exchange markets, Finance and Stochastics 1, 95-130.

Kaizoji, T. and M. Kaizoji, 2003, Empirical laws of a stock price index and a stochastic model, Advances in Complex Systems 6 (3) 1-10.

Kaizoji, T. 2004a, Inflation and Deflation in financial markets, Physica A in press.

Kaizoji, T. 2004b, Modeling of return distributions, mimeo.

Laherrere J. and D. Sornette, 1999, Stretched exponential distributions in nature and economy: Fat tails with characteristic scales, European Physical Journal B 2, 525-539.

Liu, Y., P. Gopikrishnan,, P. Cizeau,, M. Meyer, C-K. Peng, and H. E. Stanley, 1990, Statistical properties of the volatility of price fluctuations, Physical Review E 60 (2) 1390-1400.

Longin F.M., 1996, The asymptotic distribution of extreme stock market returns, Journal of Business 96, 383-408.

Lux T., 1996, The stable Paretian hypothesis and the frequency of large returns: an examination of major German stocks, Applied Financial Economics 6, 463-475.

Mandelbrot B., 1963, The variation of certain speculative prices, Journal of Business 36, 392-417.

Muller U.A., M.M.Dacarogna, O.V.Picktet, 1998, Heavy Tails in High-Frequency Financial Data, In: A Practical Guide to Heavy Tails, pp.55-78, Eds. R.J.Adler, R.E.Feldman, M.S.Taqqu, Birkhauser, Boston.

Pagan A., 1996, The econometrics of financial markets, Journal of Empirical Finance 3, 15-102.

Plerou, V., P. Gopikrishnan, L. A. N. Amaral, M. Meyer, and H. E. Stanley, Scaling of the distribution of price fluctuations of individual companies, Phys. Rev. E 60, 6519-6529.

Vries, de C.G., 1994, Stylized facts of nominal exchange rate returns, in The Handbook of International Macroeconomics, F. van der Ploeg (ed.), 348-389 (Blackwell).

------------------------------
**Figure Captions**
------------------------------

Figure 1: The movements of the stock price indices: (a) Nikkei 225 (b) DJIA, (c) SP500, (d) FT100

Figure 2: Comparisons of the complementary cumulative distribution of absolute log returns *V* for each of the four stock price indices in the period of booms with that in the period of stagnations. The dark blue circles denote the distributions in the period of booms, and the pink triangles the distribution in the period of stagnations. The distributions are shown in a semi-log scale.

Figure 3: The panels (a), (c), (e) and (g) indicate the complementary cumulative distribution of absolute log returns *V* for each of the four stock price indices in the period of booms, and the panels (b), (d), (f) and (h) indicate that in the period of stagnations. These figures are shown in a semi-log scale. The solid lines represent fits of the exponential distribution estimated from the data using a least squared method.

Figure 4: The panels (a), (b), (c) and (d) indicate the complementary cumulative distribution of absolute log returns *V* for each of the four stock price indices in the period of booms in a log-log scale. The solid lines represent the best fits of the exponential distribution estimated from the data in the large value of V using a least squared method.

Figure 5: The panels (a) and (b) show the complementary cumulative distributions of V for the four indices in the period of booms in a semi-log scale, and those in the period of stagnations in a log-log scale.

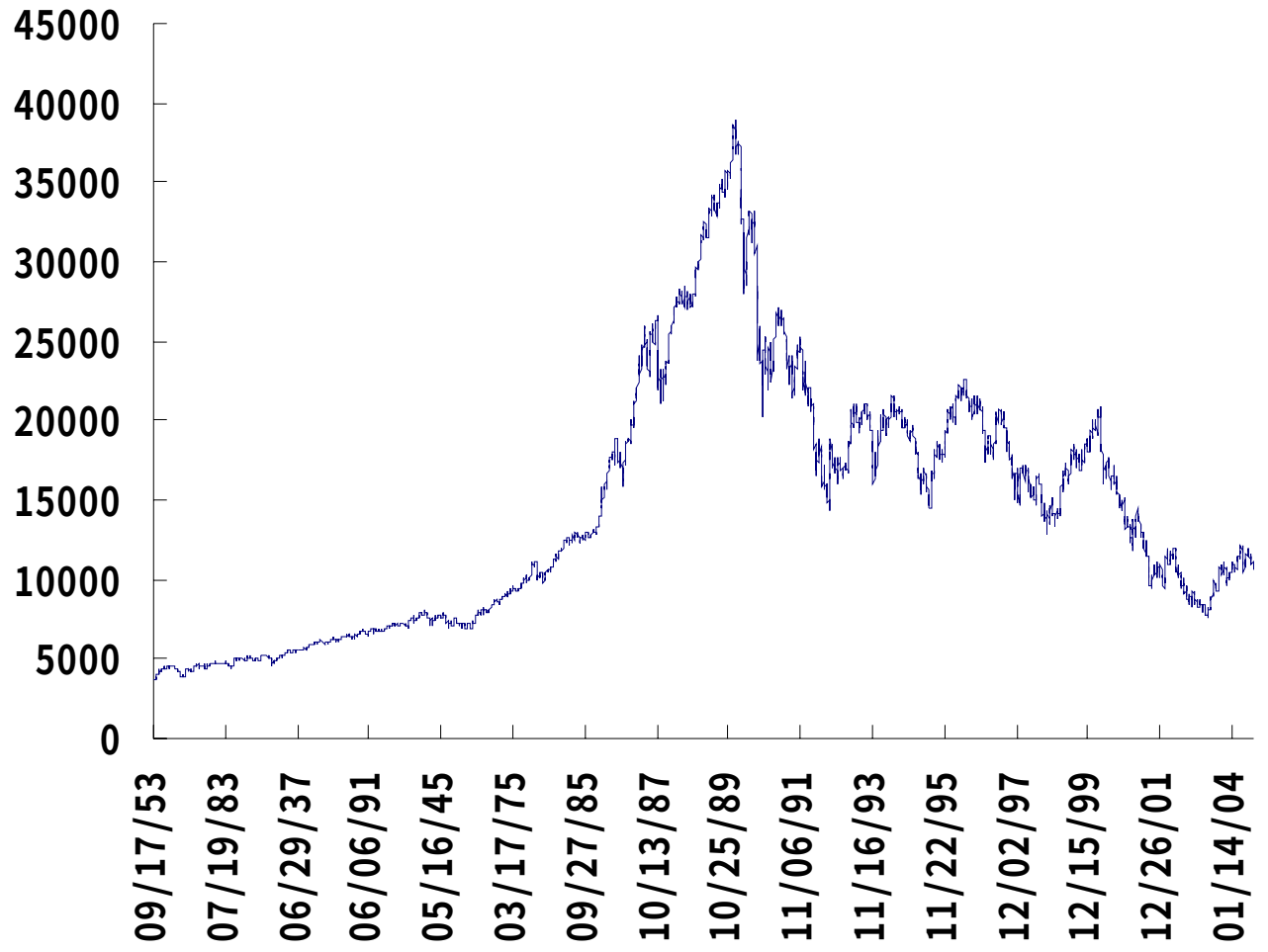

(a) Nikkei 225

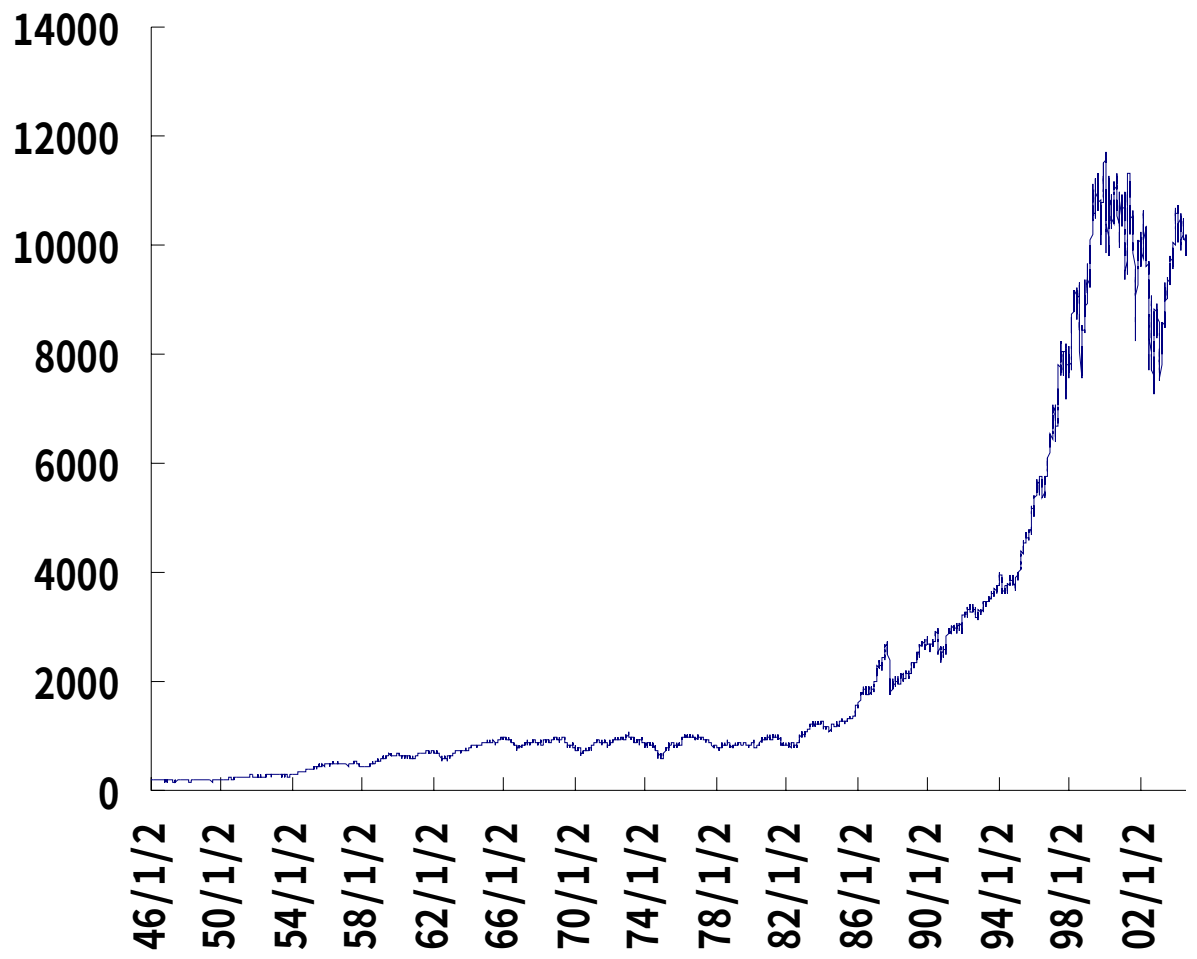

(b) DJIA

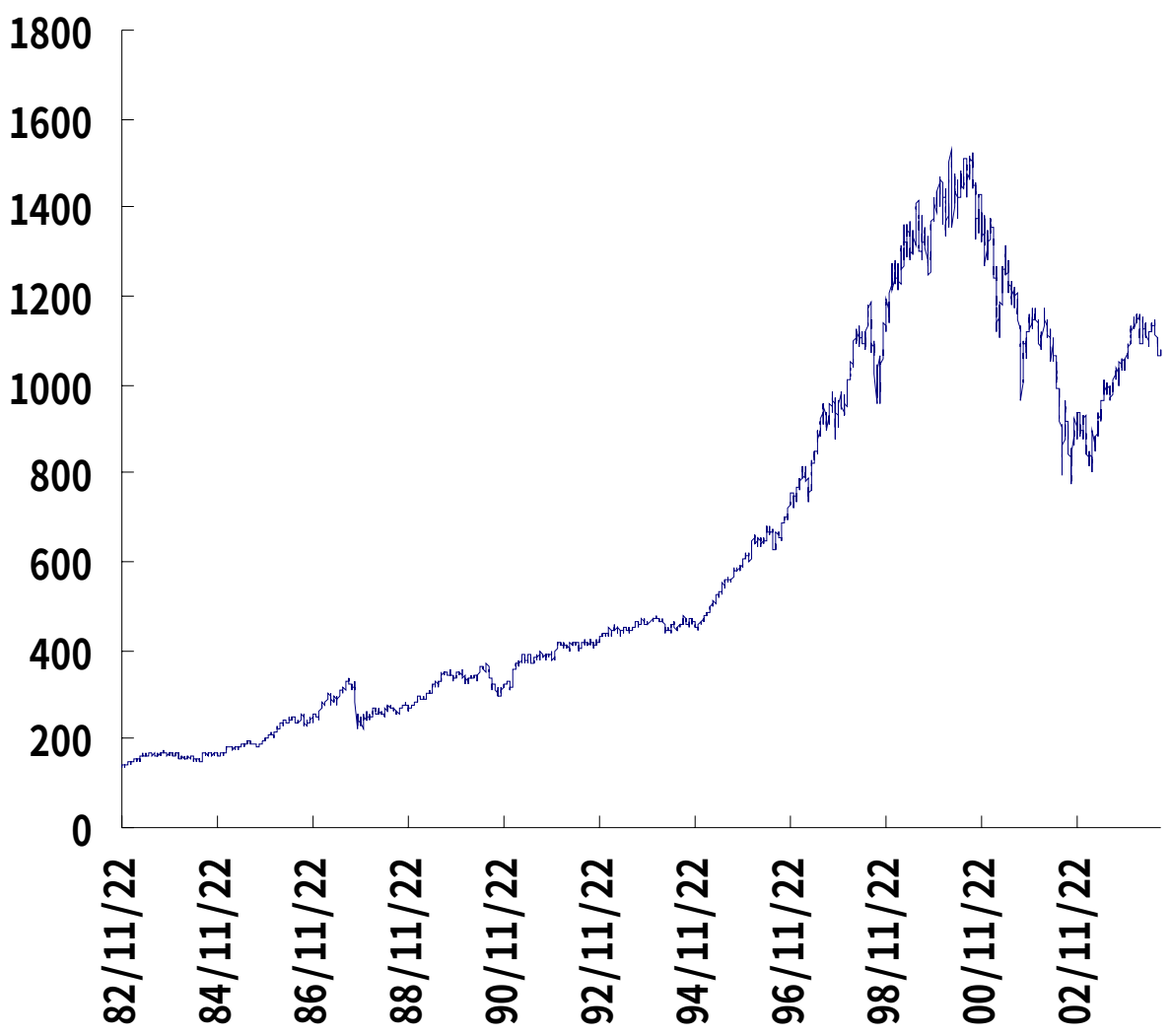

(c) SP500

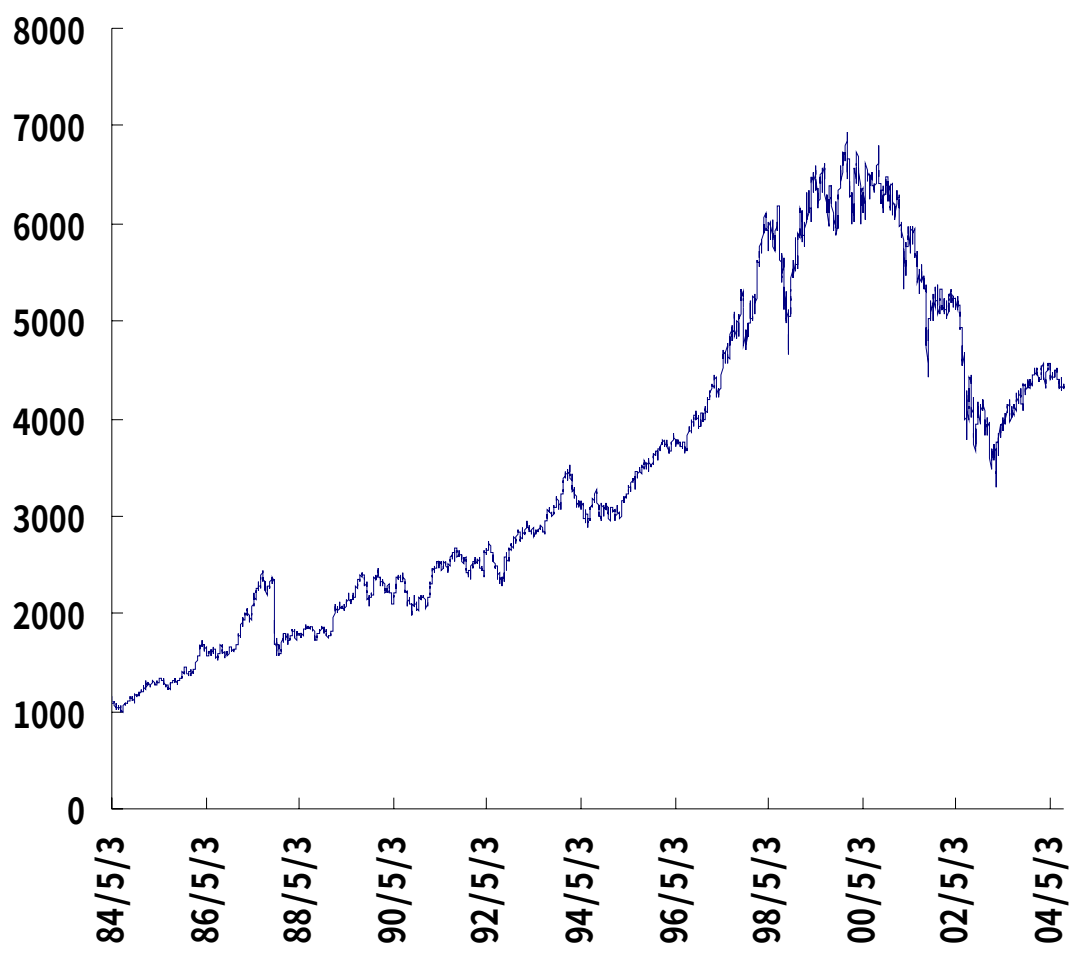

(c) FT100

Figure 1

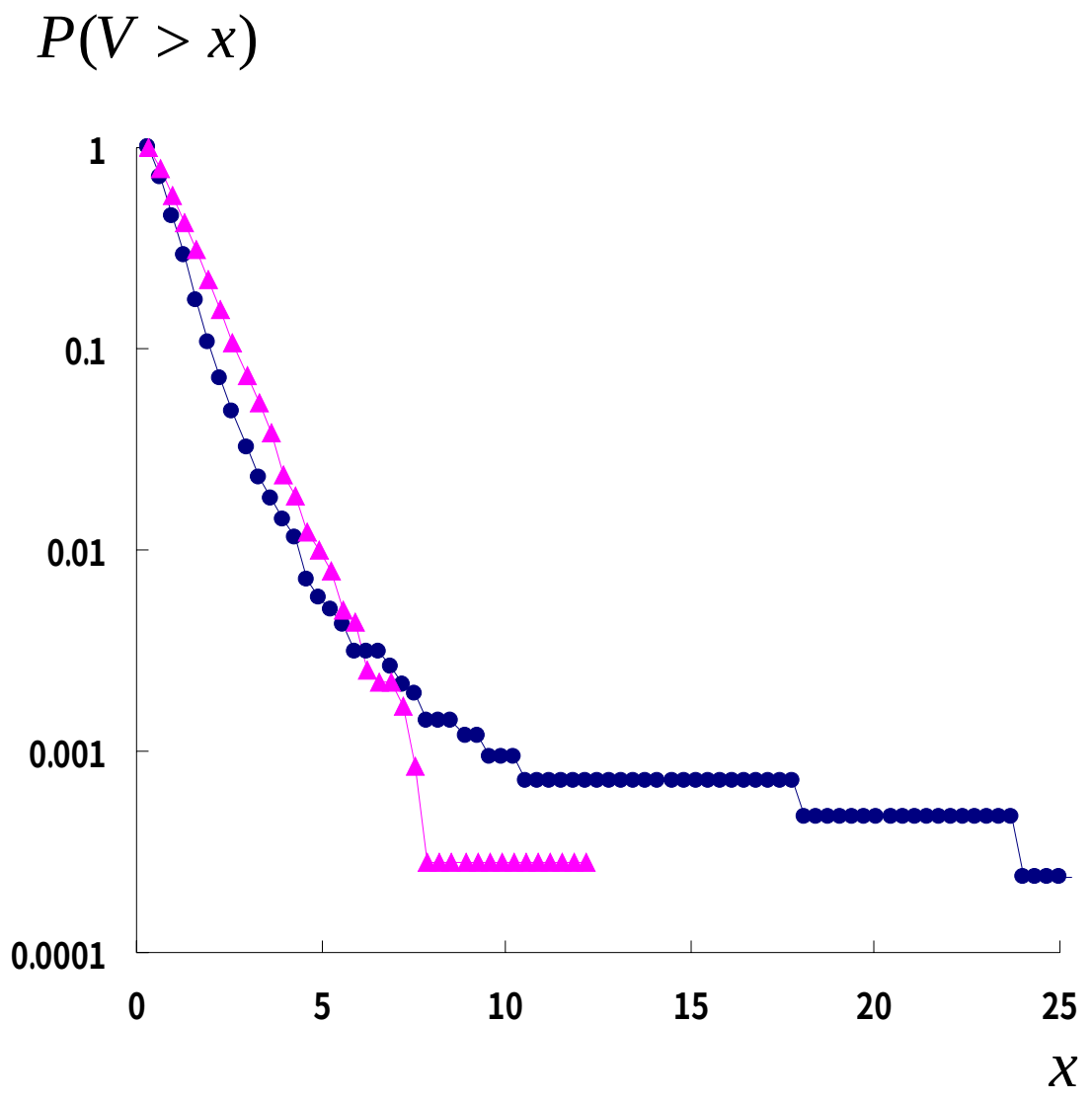

(a) Nikkei225

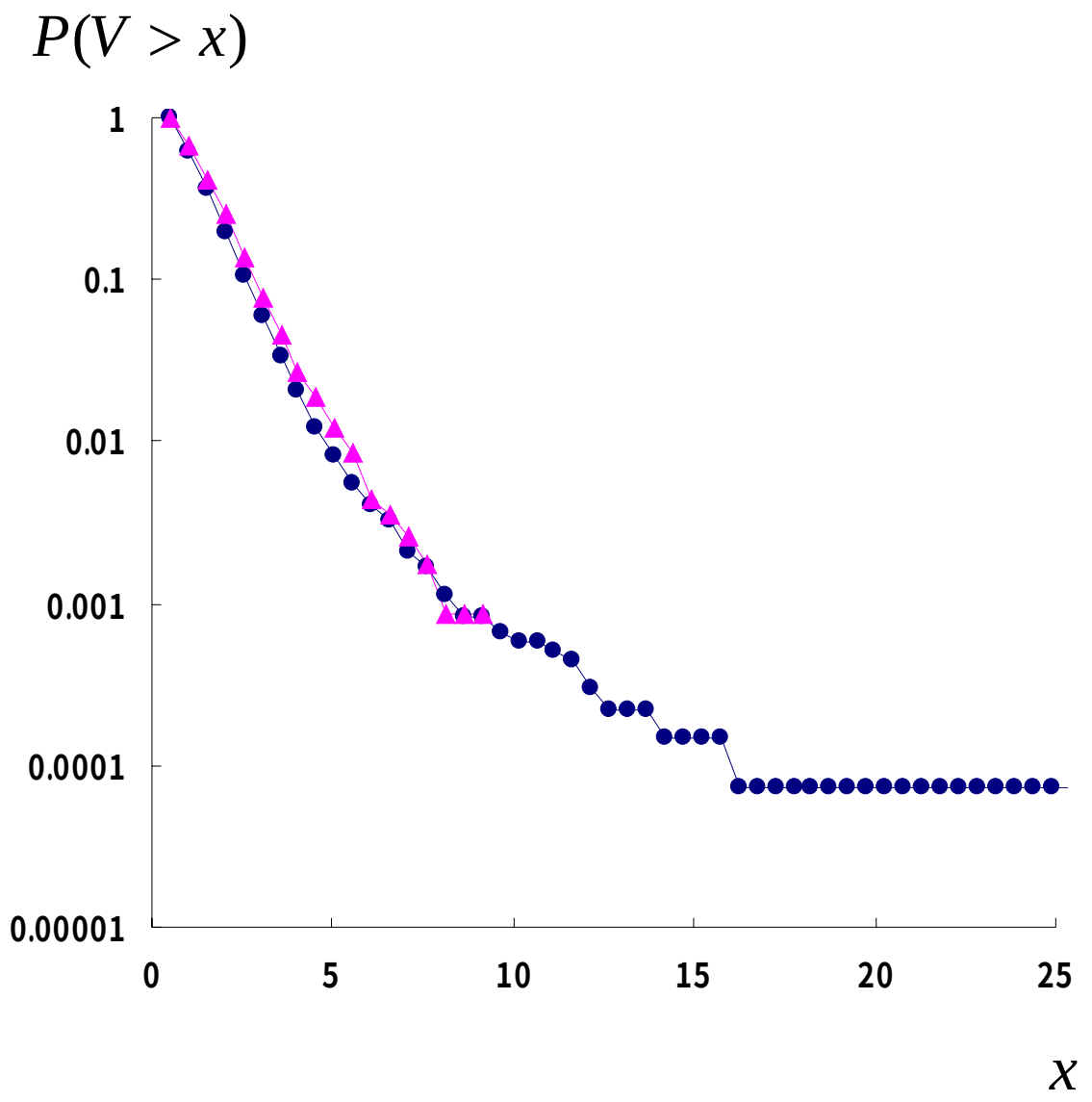

(b) DJIA

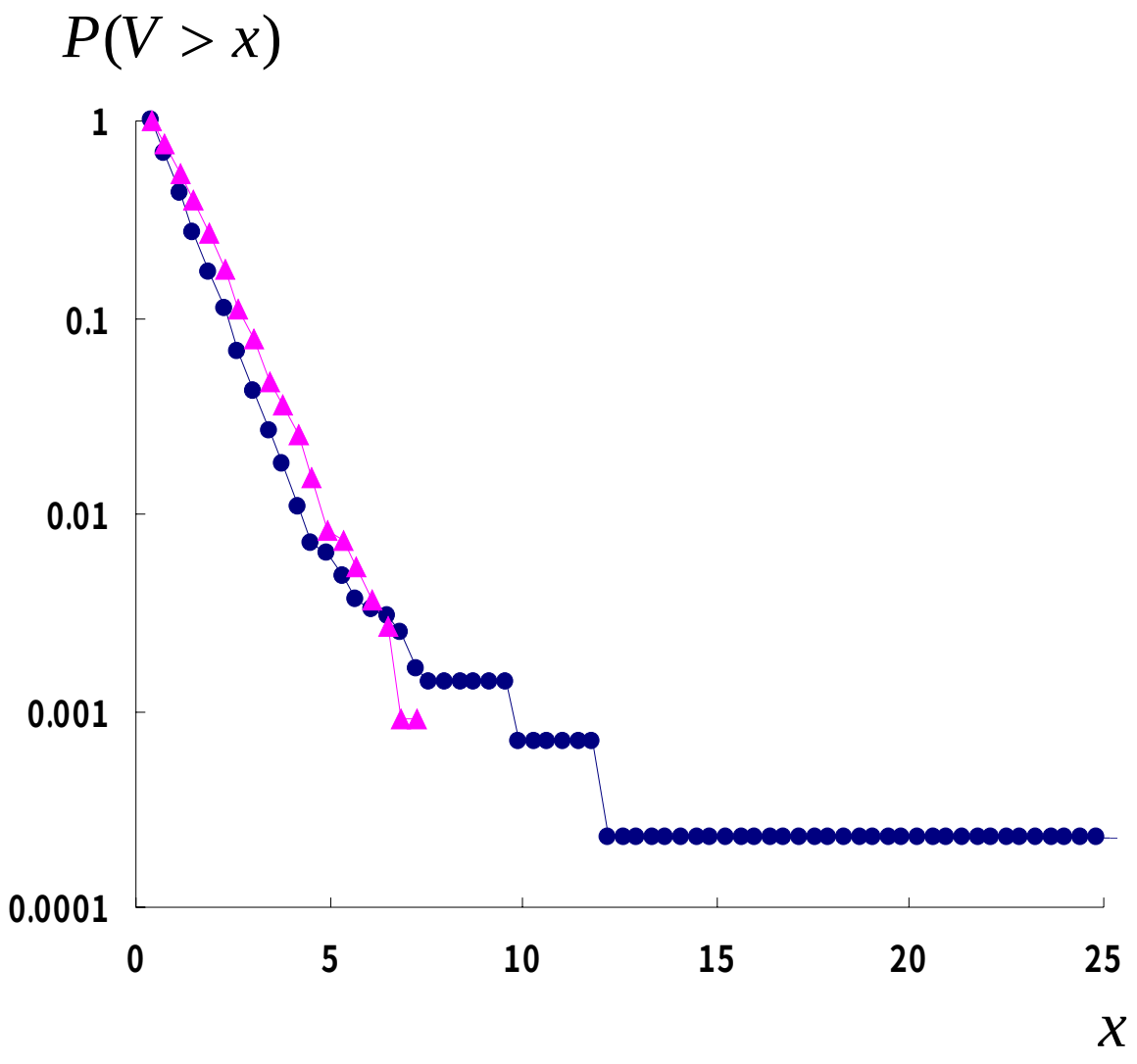

(c) SP500

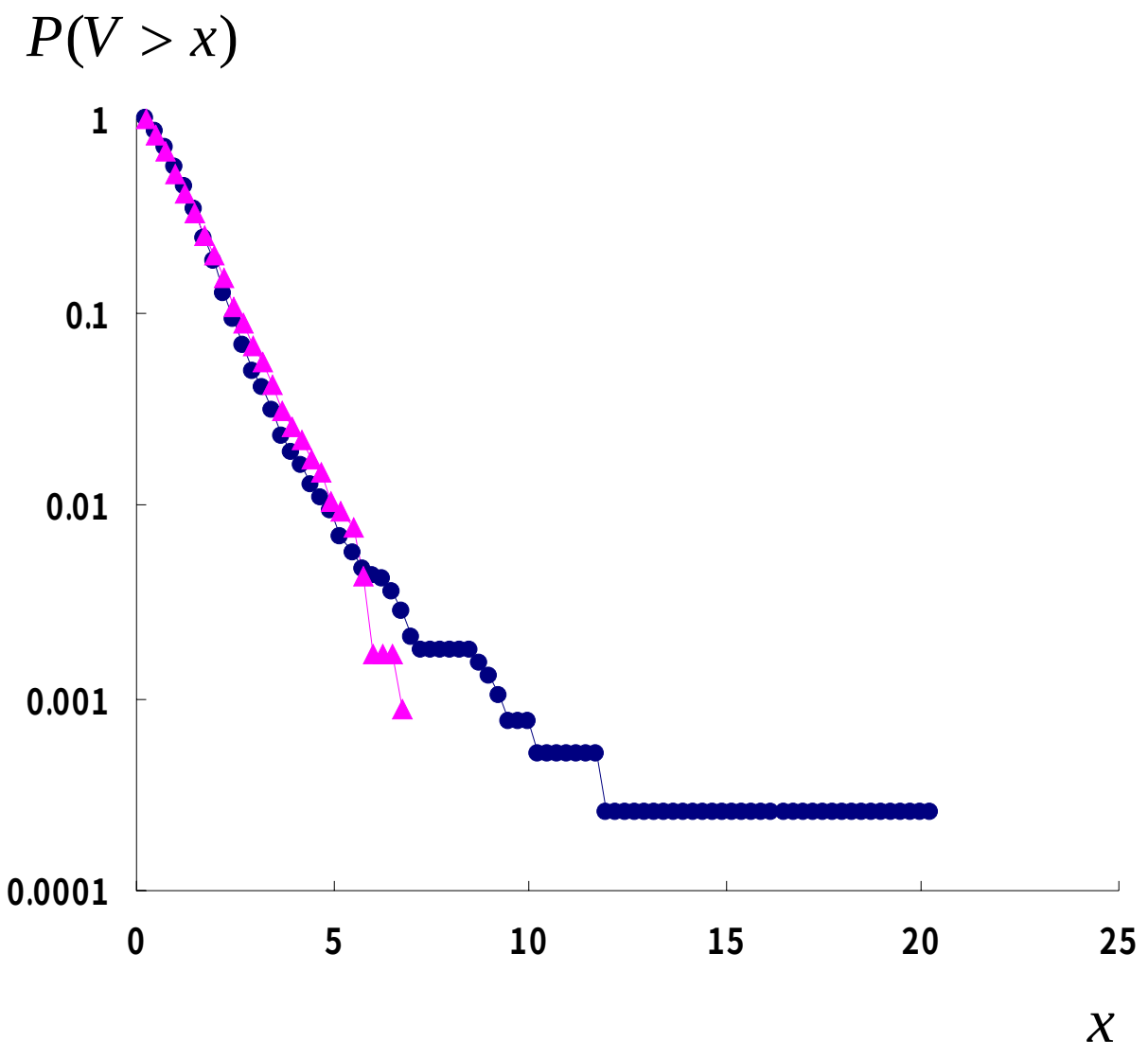

(d) FT100

Figure 2

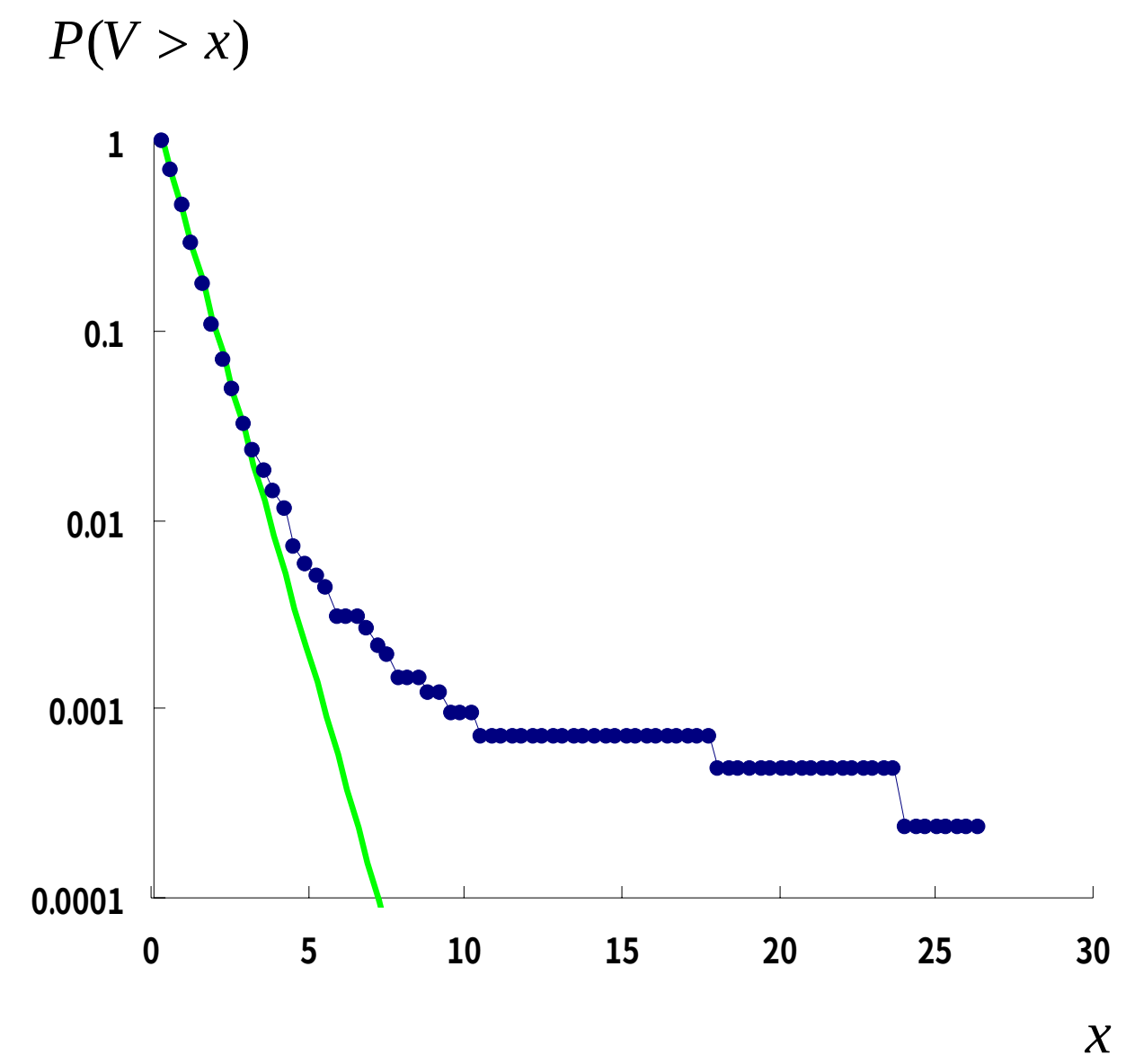

(a)     Booms in Nikkei225

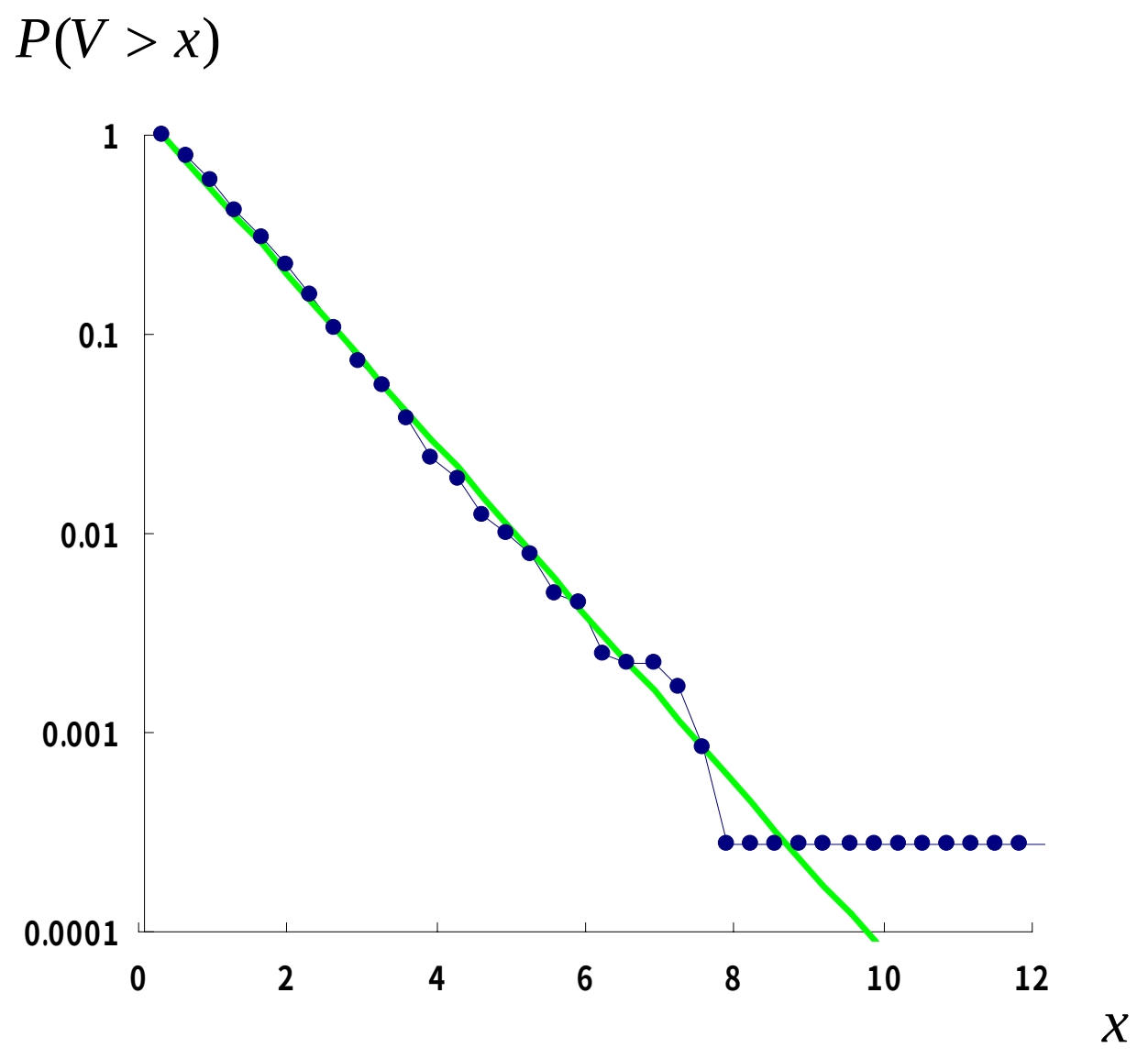

(b) Stagnations in Nikkei225

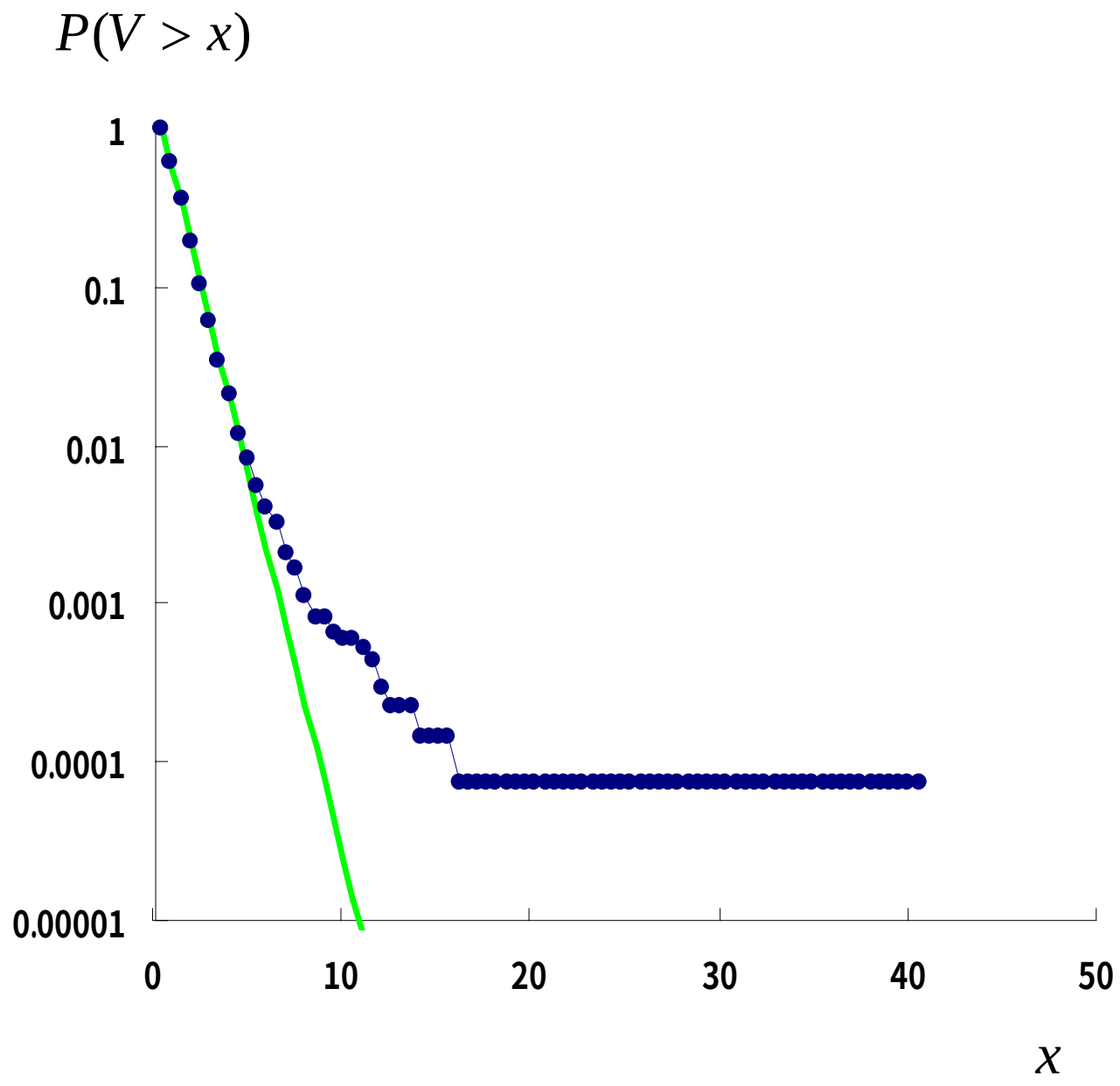

(c) Booms in DJIA

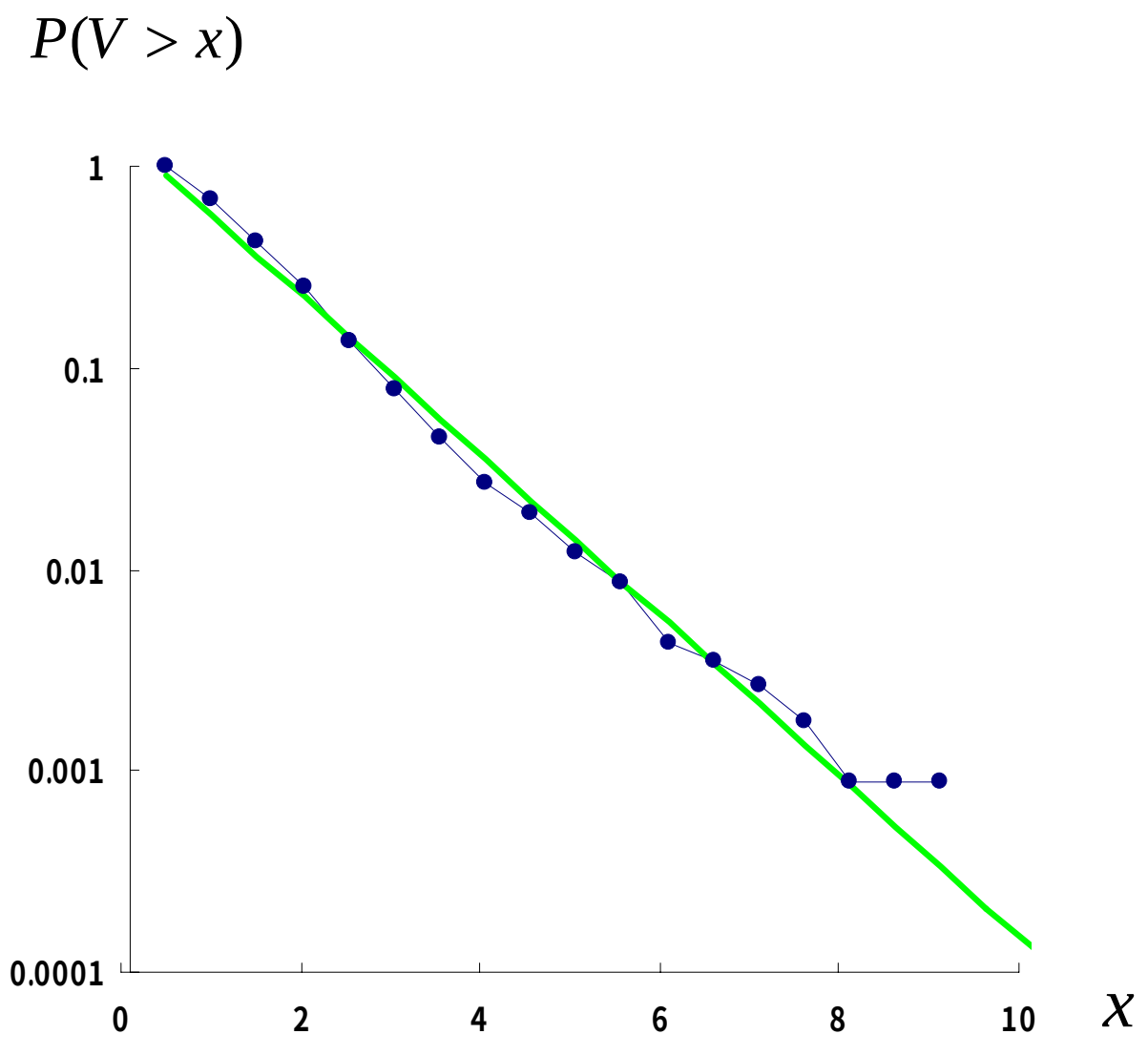

(c) Stagnations in DJIA

# Figure 3

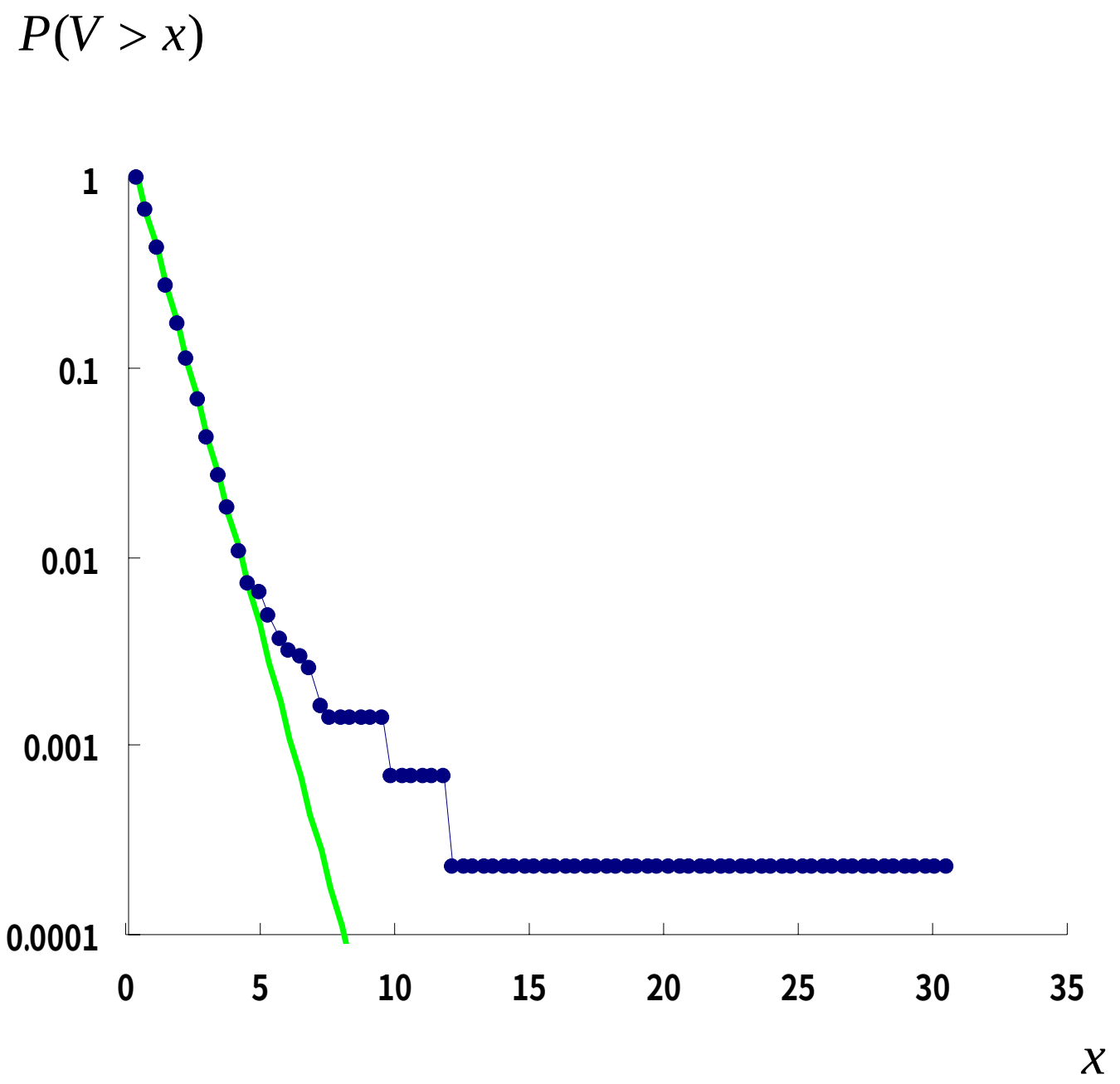

(e) Booms in SP500

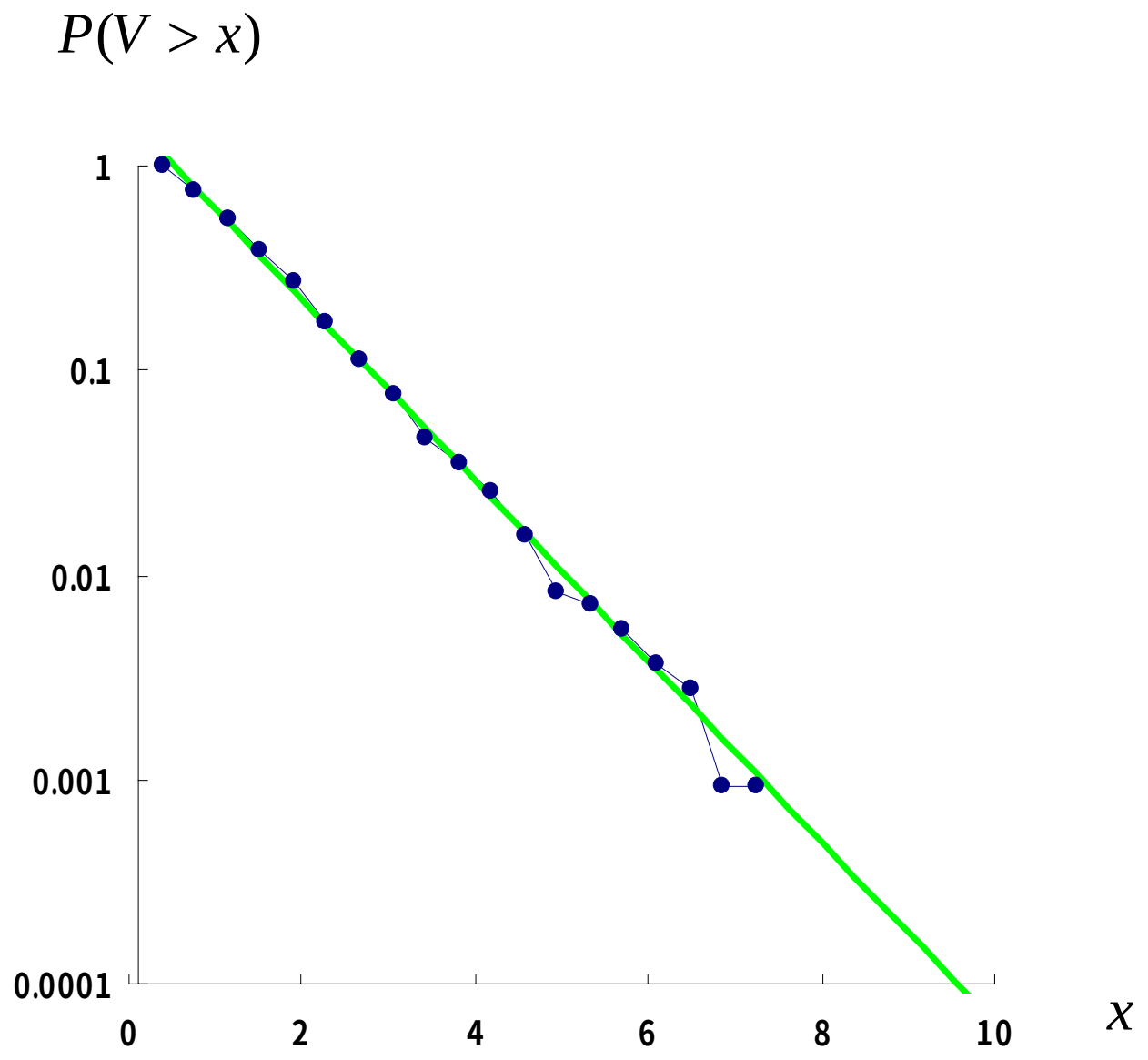

(f) Stagnations in SP500

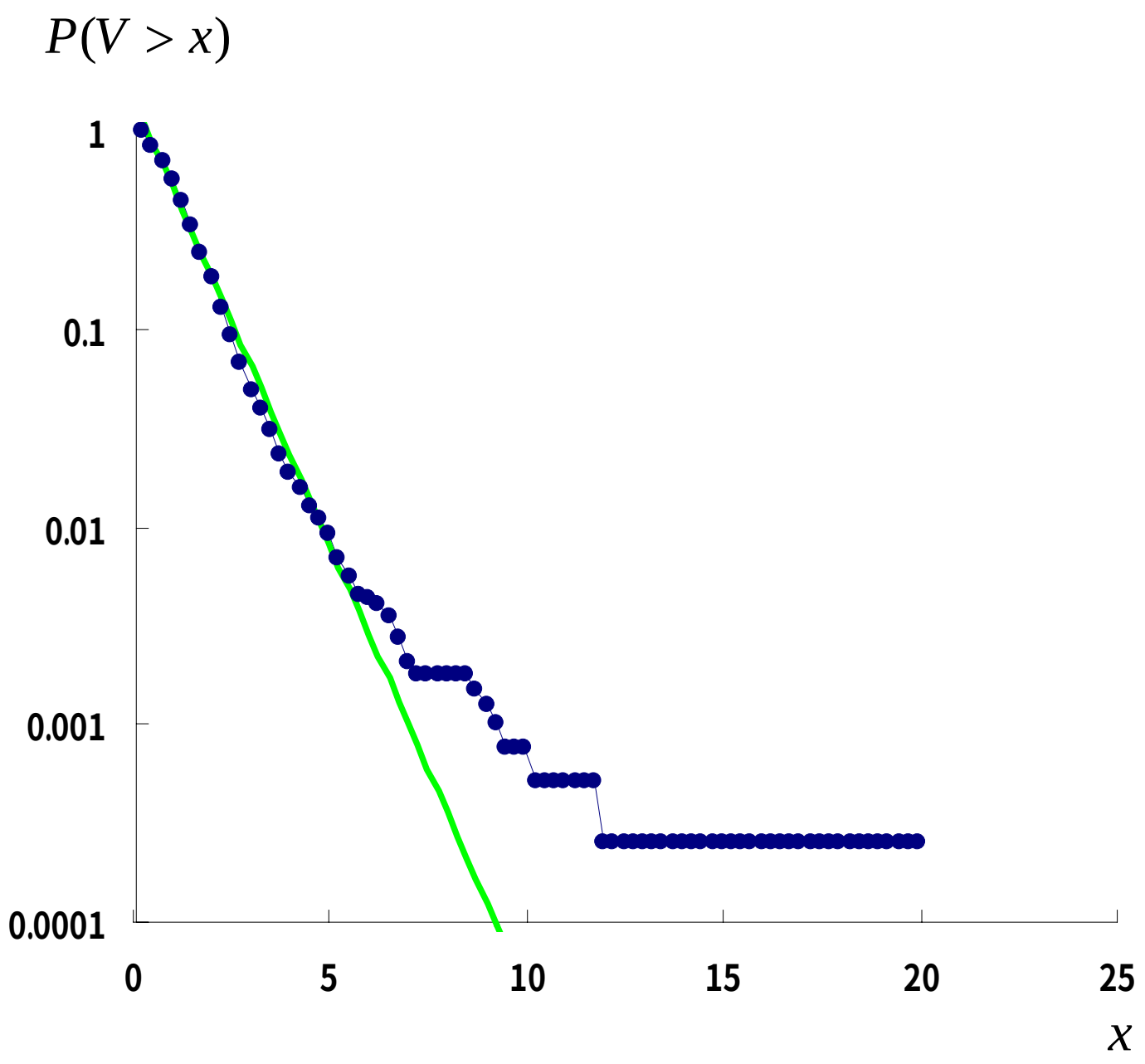

(g) Booms in FT100

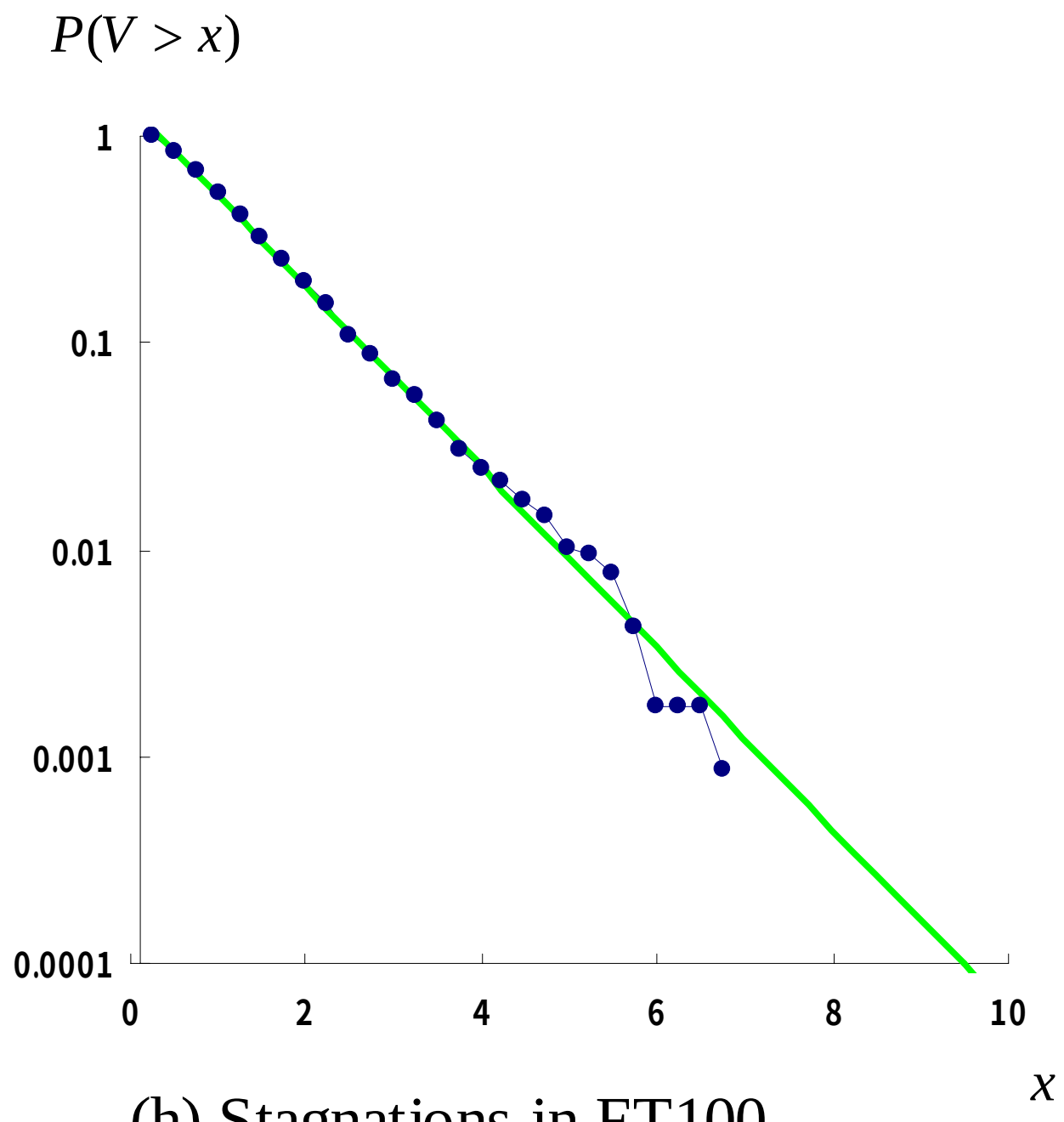

(h) Stagnations in FT100

Figure 3 (continued)

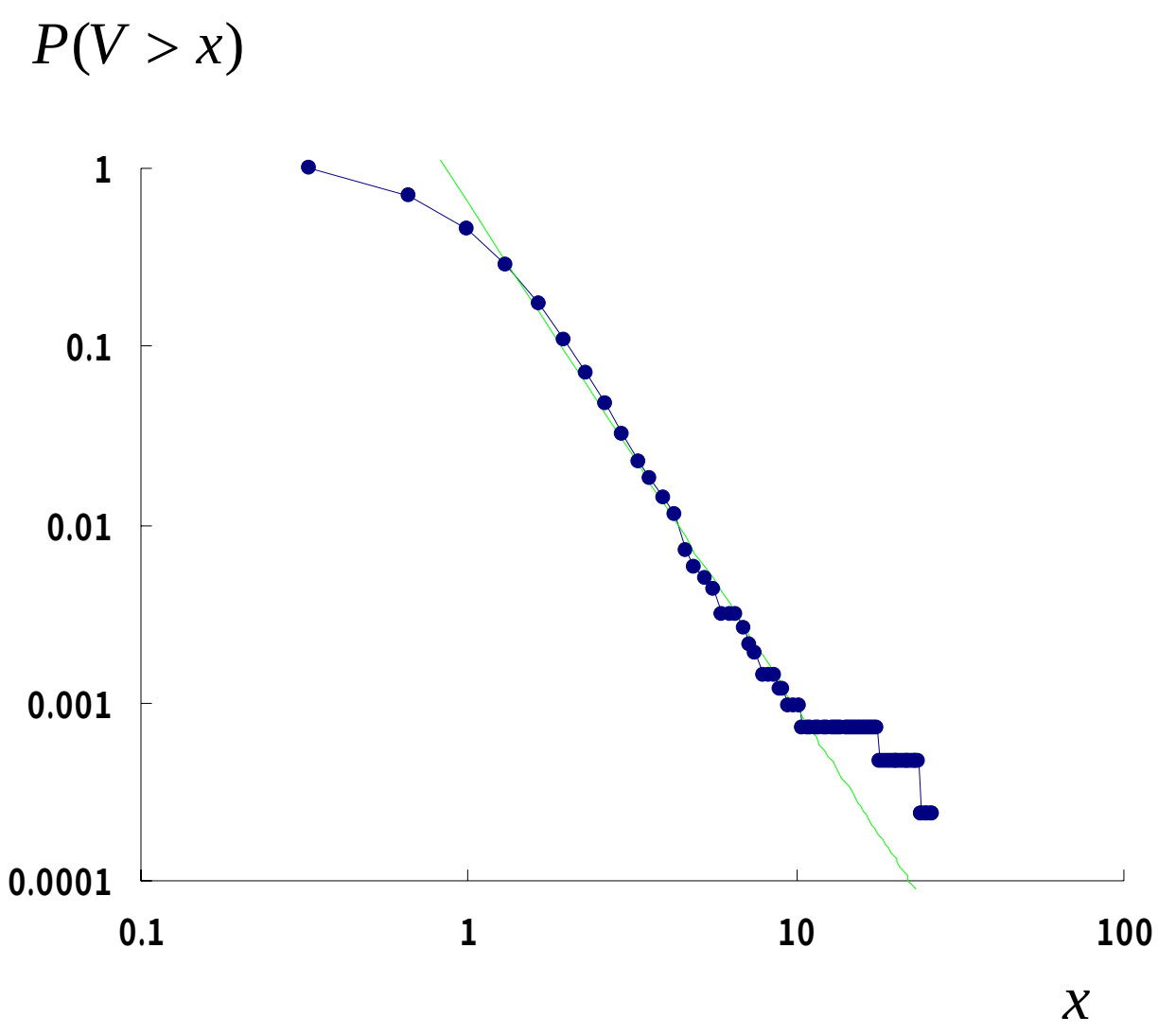

(a) Nikkei225

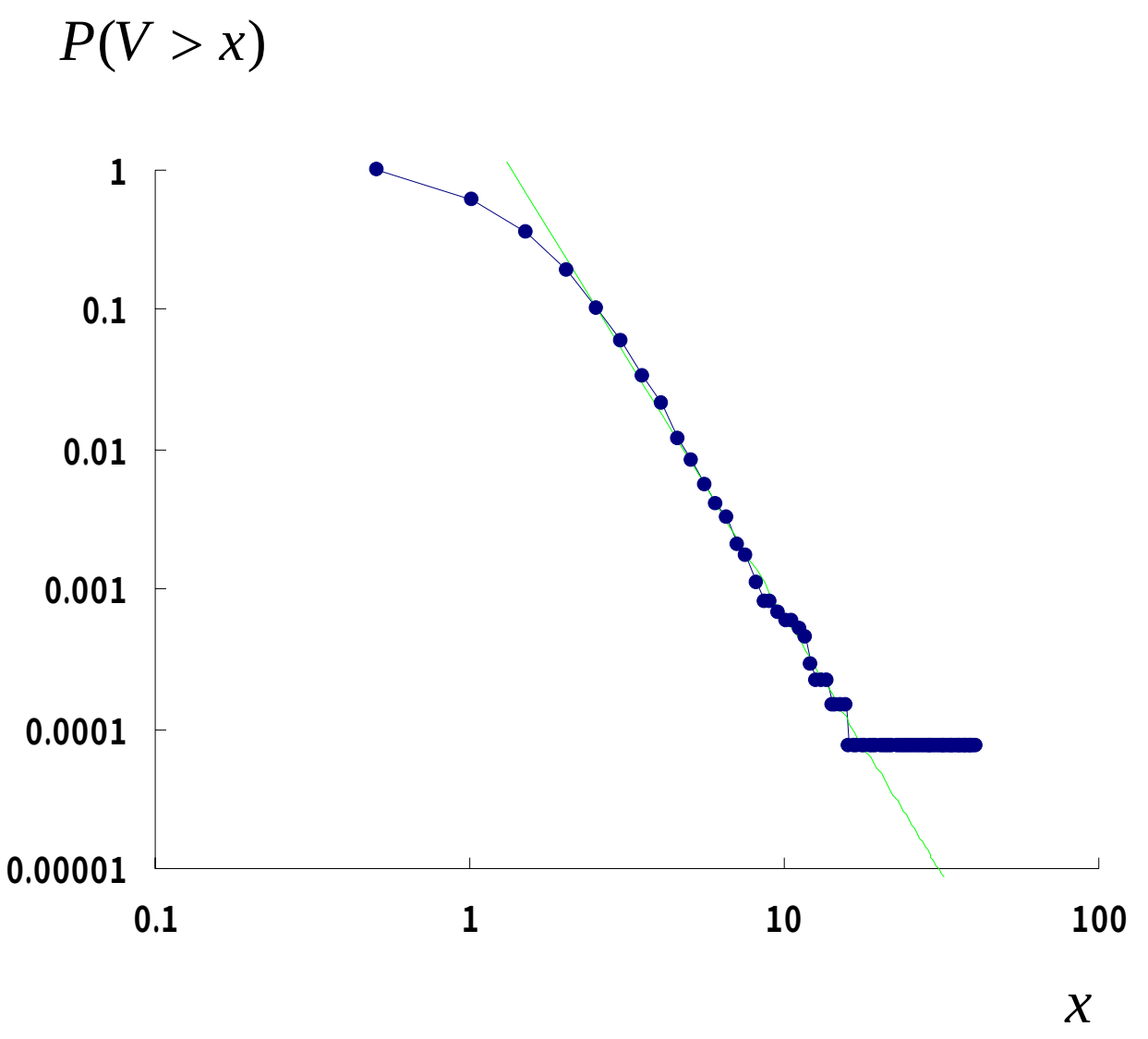

(b) DJIA

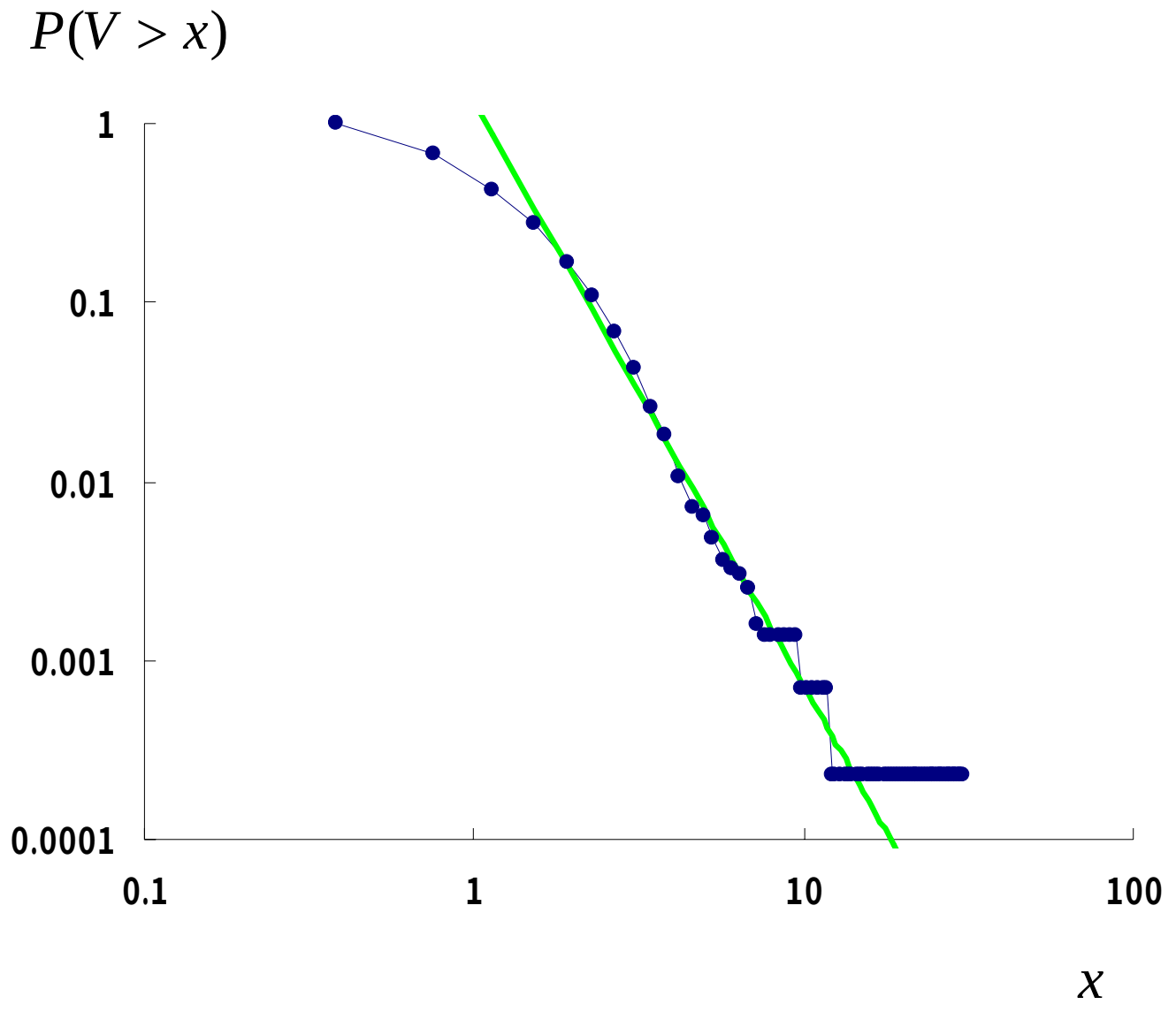

(c) SP500

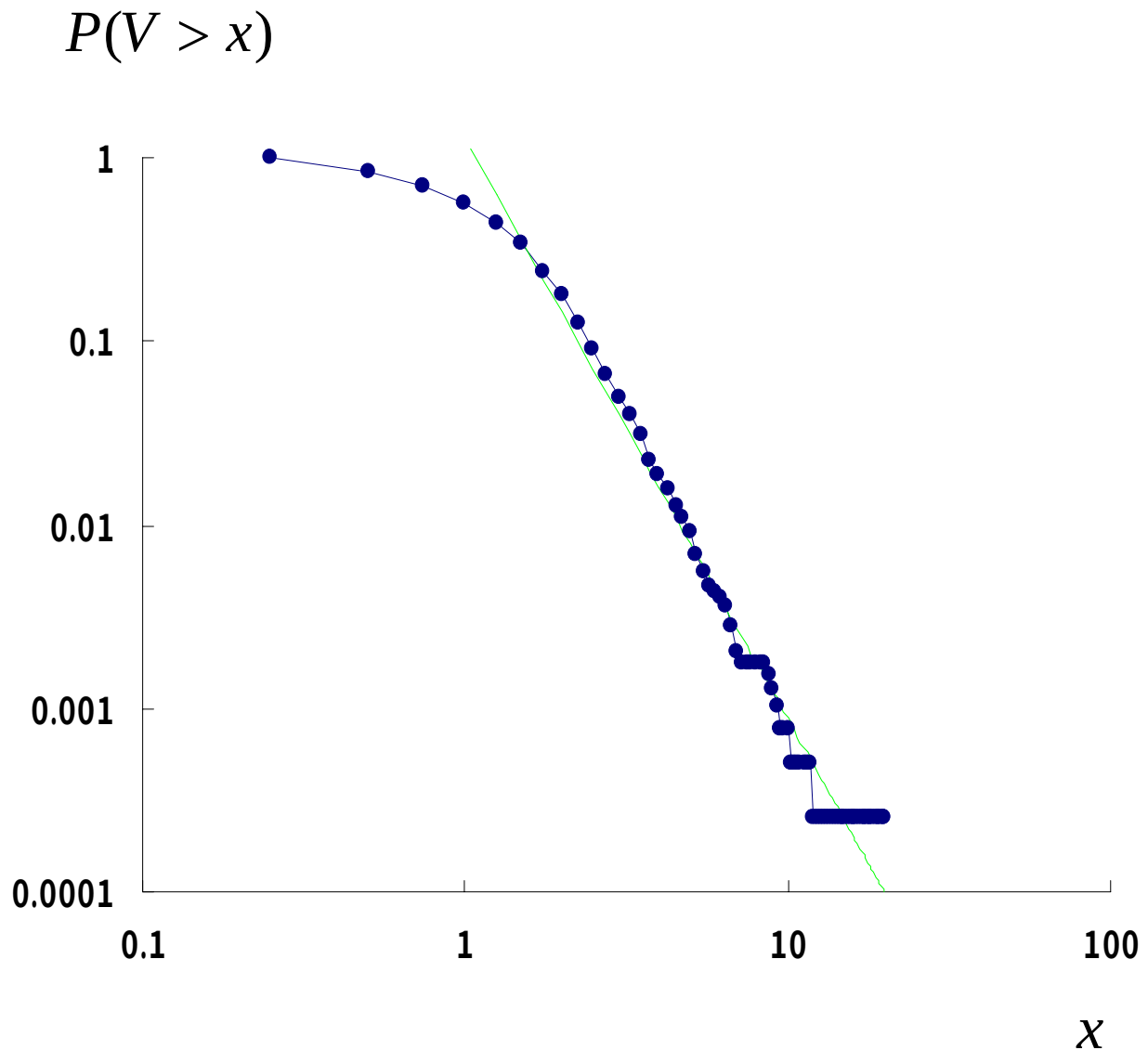

(d) FT100

Figure 4

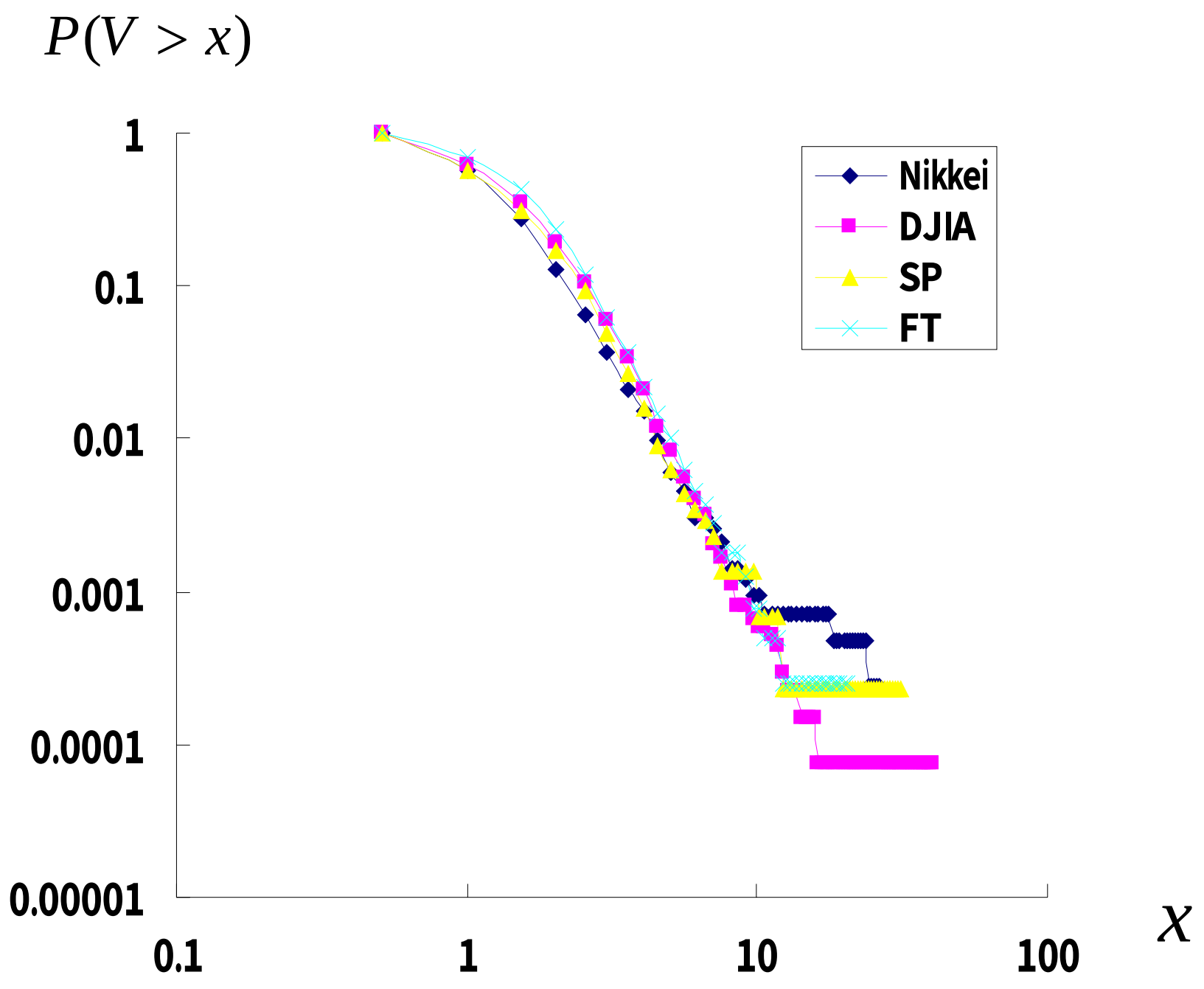

(a) Booms

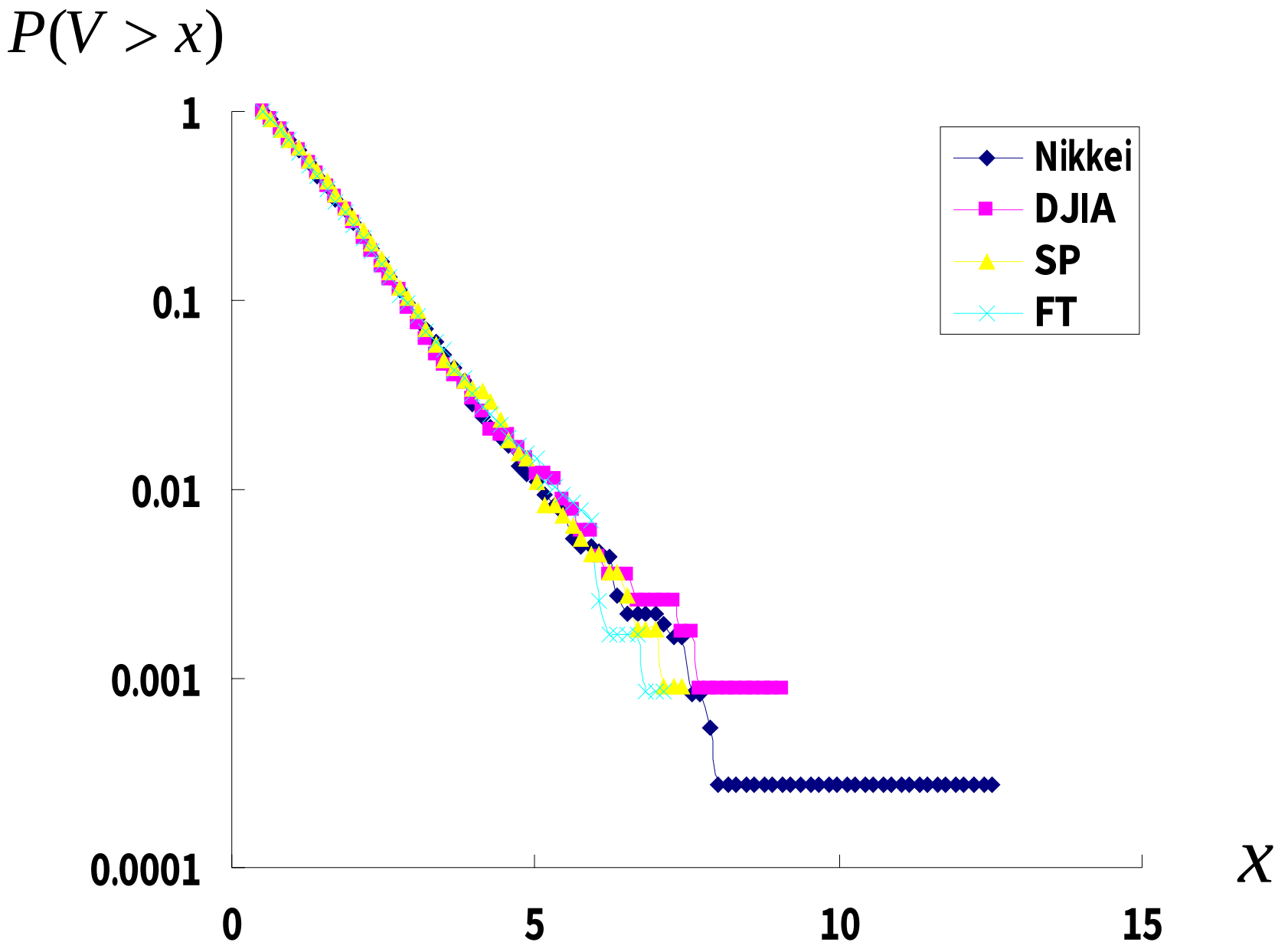

(b) Stagnations

Figure 5